\documentclass[usenatbib,usegraphicx]{mn2e}

\usepackage[english]{babel}
\graphicspath{{figures/}}

\begin{document}

\title[A huge flare on the M9 dwarf 2MASSW~J1707183+643933]{Multiband photometric detection of a huge flare on the M9 dwarf 2MASSW~J1707183+643933}
\author[Rockenfeller et al.]{Boris~Rockenfeller,$^1$
  Coryn~A.~L.~Bailer-Jones,$^1$ Reinhard~Mundt,$^1$ and
  \newauthor
  Mansur~A.~Ibrahimov$^2$\\
  $^1$Max-Planck-Institut f\"ur Astronomie, K\"onigstuhl 17, 69117
  Heidelberg, Germany\\
  $^2$Ulugh Beg Astronomical Institute, 33 Astronomical Str., Tashkent
  700052, Uzbekistan}

\date{Accepted 2005 December 06.  Received 2005 December 05; in
original form 2005 October 31}

\maketitle

\begin{abstract}
We present simultaneous UV-G-R-I monitoring of 19 M~dwarfs that
revealed a huge flare on the M9~dwarf %\object{2MASSW~J1707183+643933}
with an amplitude in the UV of at least 6 magnitudes. This is one of
the strongest detections ever of an optical flare on an M~star and
one of the first in an ultracool dwarf (UCD, spectral types later than
about M7). Four intermediate strength flares ($\Delta m_{\rm
UV}<4\,{\rm mag}$) were found in this and three other targets. For
the whole sample we deduce a flare probability of $0.013$ (rate
of $0.018\,{\rm hr}^{-1}$), and $0.049$ ($0.090\,{\rm hr}^{-1}$)
for 2M1707+64 alone. Deviations of the flare emission from a blackbody
is consistent with strong $H_{\alpha}$ line emission. We also confirm
the previously found rotation period for 2M1707+64 \citep{Rockenfeller2006a}
and determine it more precisely to be $3.619\pm 0.015\,{\rm hr}$. 
\end{abstract}


\section{Introduction}
Numerous early and mid M~dwarfs are known to show short timescale
brightness eruptions -- flares similar to those on the sun. They are
probably caused by magnetic fields, whereby energy is locally
converted to thermal energy by a reconnection of field lines, the
consequence of which is a significant increase in the total stellar
luminosity. Flares manifest themselves in various wavelength regions,
from the radio to the X-ray and as line or continuum emission. The
corresponding motions of plasma have been detected by emission line
asymmetry \citep[e.g.]{Fuhrmeister2005}. In the recent years, a few
flares in the optical as well as emission line flares have also been
observed in late~M and L~dwarfs (see below).

The creation process of magnetic fields in fully convective objects
(such as UCDs) is not yet fully understood. But it must differ from
the $\alpha\Omega$-dynamo present in higher mass stars because this
depends on the interface between radiative and convective
zones. However, the detection of flares as well as cool surface spots
strongly suggests the existence of magnetic fields in UCDs. Recently,
Chabrier \& K\"uker (astro-ph/0510075) modelled an $\alpha^{2}$-dynamo
for ultracool dwarfs. Among other things, they are able to explain the
saturation in the rotation--activity relation which has been observed
in these objects \citep{Delfosse1998,Mohanty2003}.

Amplitudes of flares on K and M~dwarfs in the optical have
been observed to increase strongly from the red (I-band) to the U/UV
\citep{Stepanov1995,Eason1992} and no signatures can be traced in the
near-infrared. The duration of these events seems to positively
correlate with their amplitude and ranges from seconds to a
few hours.

Although several extensive photometric monitoring programs of L and
T~dwarfs have been carried out, only a few flares have been reported,
probably because most of these surveys were conducted in the I-band or
near-infrared~(JHK). Nevertheless, flares have recently been
detected by optical broadband photometry, for example by
\citet{Koen2005a} in an M8.5 dwarf with an R-band amplitude of almost
2 mag and by \citet{Scholz2005} in a young VLMS. \citet{Koen2005b}
reports of a possible flare event in an object as late as L8 (R-band
amplitude of 0.15 mag). Emission line flares have also been reported,
e.g.\ by \citet{Liebert1999}~(M9.5 object), \citet{Liebert2003}~(L5)
and by \citet{Fuhrmeister2004}~(M9).

\citet{Rockenfeller2006a} (henceforth R06) found periodic variability
in the GRI-bands of 2M1707+64 at a period of $3.65\pm 0.1\,{\rm hr}$
with an amplitude of $0.014\,{\rm mag}$ in the I-band.  By modelling
the amplitudes they showed the modulation was best explained by
localised cool surface spots in this late-type object (which has a
spectral type and estimated age placing it near to the hydrogen
burning limit). $H_{\alpha}$ quiescent emission was found by
\citet{Gizis2000} at an equivalent width of $9.8\AA$, which suggests
2M1707+64 is a very active M~dwarf, even during its non-flaring state.

Here, we report the detection of two flares on 2M1707+64 of which one
has exceptionally large amplitudes throughout the optical, with a peak
of more than 6 magnitudes in the UV. We also report the detection of
smaller (but still large) flares on three other late M dwarfs.

%%% Local Variables: 
%%% mode: latex
%%% TeX-master: "Paper"
%%% End: 

\section{Data acquisition, reduction and analysis}

\begin{figure}
  \resizebox{\hsize}{!}{\includegraphics{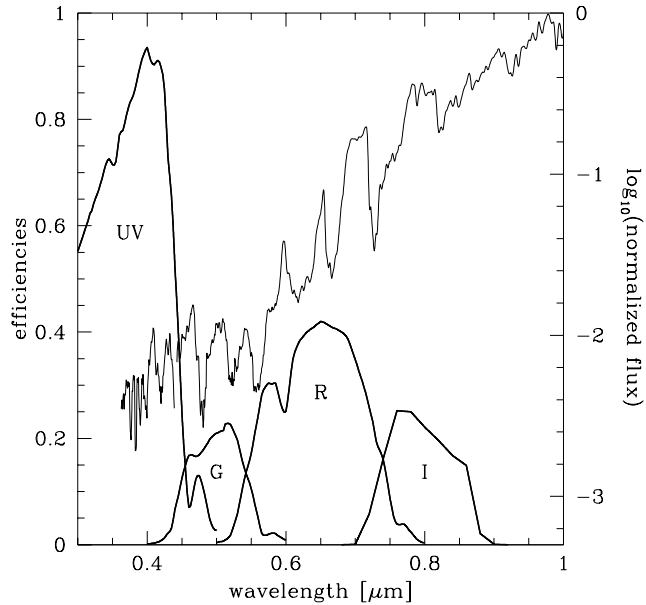}}
  \caption{The total efficiencies of the BUSCA UV, G, R and I
    channels, taking into account the dichroics, the CCD efficiency and
    the Bessell-I filter used in the I-band. The efficiency of
    the UV filter drops off rapidly blueward of $0.3\,{\rm \mu m}$ due
    to the atmosphere.  Over plotted for reference is a spectrum of
    the M4.5~dwarf DENIS$\,$P-J1158-1201 from \citet{Martin1999}.}
  \label{figure_BUSCAwavebands}
\end{figure}

Simultaneous multichannel UV,G,R,I observations of 19 M~dwarfs
were obtained over 20~nights in June 2002 and 2003 with the four
channel CCD camera BUSCA at the 2.2m telescope of the Calar Alto
Observatory, Spain. BUSCA uses dichroics to split the light beam into
the four wavelength bands which are imaged by four separate CCD
cameras.  Fig.~\ref{figure_BUSCAwavebands} shows the passbands. The
UV, G and R passbands are defined by the CCD response and dichroic
transmission function; for the I band we additionally used a Bessel I
filter.

Follow-up observations of 2M1707+64 were performed in June 2005 on two
nights (14th and 15th) -- again in the four channels with BUSCA --
during Directors Discretionary Time, as well as on three nights (24th,
26th and 29th June 2005) in the I-band with the $1.5\,{\rm m}$
telescope at Maidanak Observatory, Uzbekistan. All data were reduced
in the same way as described in R06, except that the Maidanak data
required an additional fringe subtraction. As in R06, light curves
were obtained from differential photometry using a set of stable
reference stars.  Table~\ref{table_ObservationLog} shows the
observation log. See R06 for detailed information on the analysis of
the 2002/03 BUSCA data, in particular time series analysis,
determination of rotation periods and spot modelling.
\begin{table*}
\begin{minipage}{124mm}
\caption{Observation log of the four M~flare~dwarfs, listing the
  objects' names, the dates of observations, the duration of observation the individual
  nights (in hours) and the instrument/site.}
\label{table_ObservationLog}
\begin{tabular}{llll}
\hline
object  &  dates  &  duration [hr]  &  instrument\\
\hline
2MASSW J1707183+643933  &  2002.06.01 / 02         &  6.4, 5.6  &  BUSCA\\
                        &  2005.06.14 / 15         &  5.7, 3.9  &  BUSCA\\
                        &  2005.06.24 / 26 / 29    &  5.6, 5.1, 4.4  &  Maidanak\\
2MASSW J1344582+771551  &  2002.05.29 / 2005.06.01 &  5.2, 6.2  &  BUSCA\\
2MASSW J1546054+374946  &  2003.06.06 / 07 /08     &  6.4, 2.6, 2.4  &  BUSCA\\
2MASSW J1714523+301941  &  2003.06.03 / 05         &  6.1, 5.4  &  BUSCA\\
\hline
\end{tabular}
\end{minipage}
\end{table*}

%%% Local Variables: 
%%% mode: latex
%%% TeX-master: "Paper"
%%% End: 

\section{Flares on 2M1707+64}\label{section_Flares2M1707+64}

\begin{figure}
\includegraphics[width=84mm]{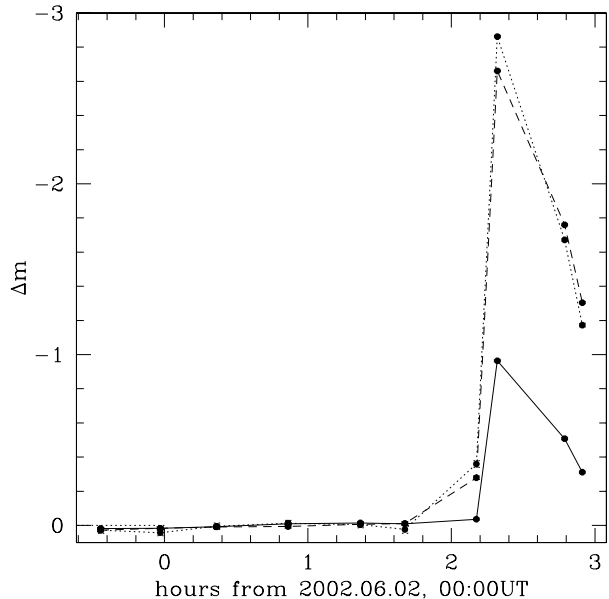}
\caption{Relative light curves of 2M1707+64 on 2002.06.02, showing the
I-band (solid line), R-band (dashed line) and G-band data (dotted
line). The UV-band flare amplitude is larger than 6 magnitudes. The
error bars are also plotted, but except for the G-band are too small
to be visible.}
\label{figure_LC2002}
\end{figure}

\begin{figure}
\includegraphics[width=84mm]{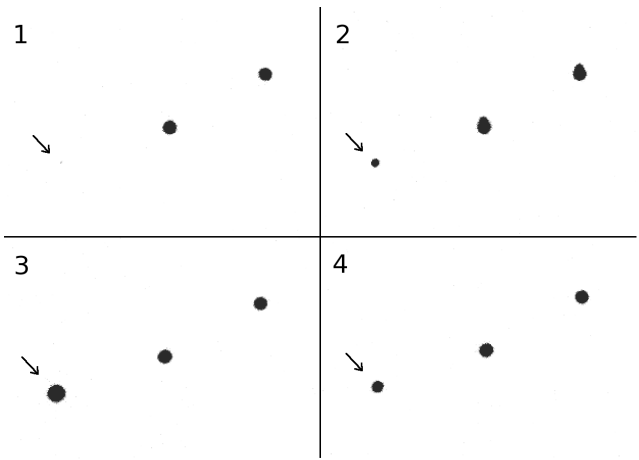}
\caption{Set of four UV-band images of 2M1707+64 on 2002.06.02. The first
image (upper left) shows the object during quiescent state, whereas the
following three show the first three (out of four) images of the flare event
-- the numbers denote the time-order. Each image is a $\approx 2.4\arcmin\cdot
1.8\arcmin$ clip from the whole field, with North up and East to the left.}
\label{figure_UVimages}
\end{figure}

The time series of 2M1707+64 shows a large increase in brightness toward the
end of observations on the morning of 2002.06.02 with increasing amplitudes
from I to UV (see Fig.~\ref{figure_LC2002}).
The wavelength dependence, the shape of the light curves and the
duration of the event (more than $1.0\,{\rm hr}$) are
consistent with the signatures of stellar flares on late-type
stars \citep{Stepanov1995}. Since the time sampling is quite coarse
with respect to the duration of the event, our observations probably
missed the maximum. Our reported amplitudes are therefore lower limits
to the maximum flare amplitudes. Furthermore, because the target is
not visible in the UV-band during its quiescent state we have
estimated a lower limit to the UV-amplitude by equating the
quiescent target magnitude to the limiting magnitude of the exposure
(which is approximately the magnitude of the faintest star visible).
The very large amplitudes in all four channels with no change in the
reference stars prove that the brightening is intrinsic to the source
and is not a result of second order extinction effects, which are far
smaller \citep{Bailer-Jones2003}. The brightness evolution in the
UV-band can be seen in Fig.~\ref{figure_UVimages}.

The flare amplitudes in the I-, R-, G- and UV-band are $1.0$, $2.7$,
$2.9$ and $6$ magnitudes, respectively, which is exceptionally large
(compare the papers mentioned in the introduction). This is one of the
strongest broadband optical flares ever reported (more so as
we probably missed the maximum of the event). As a comparison, see
\citet{Batyrshinova2001} who recorded a superflare on an M4~dwarf with
an amplitude in the U-band of $11\,{\rm mag}$.

If we initially assume that the radiation from the flare is
thermal and so obeys a blackbody distribution, we can attempt to
determine the plasma temperature ($T_{\rm p}$) and the (projected)
fractional area of the stellar surface that is effected by the flare
($A$). We used equation~1 of \citet{deJager1986} for an optically
thick plasma to simulate flare amplitudes over a wide parameter space
($5000\,{\rm K}<T_{\rm p}<10^{5}\,{\rm K}$ and
$10^{-7}<A<0.5$). Because of the low effective temperature of
2M1707+64, we used an appropriate synthetic spectra of
\citet{Allard2001} to model the flux during the quiescent state. The
goodness-of-fit of an individual model was judged by a simple function
$f(T_{\rm p},A)=1/4\cdot (\left|\delta_{\rm
UV}\right|+\left|\delta_{\rm G}\right|+\left|\delta_{\rm
R}\right|+\left|\delta_{\rm I}\right|)$, where $\delta_i$ is the
difference in magnitudes between the observed amplitude and the
simulated amplitude in channel $i$. This reveals a degeneracy in the
model (i.e.\ doubling the temperature and halving the area leads to
the same value of $f$). Moreover, the large values of $f(T_{\rm
p},A)$ for all fits indicate that the blackbody approximation for this
flare's emission is inappropriate (and/or that the quiescent spectral
model is poor). Even the best-fitting simulations could only fit 2
out of 4 channels to within $0.1\,{\rm mag}$ (UV and G-band), the
third within $0.6\,{\rm mag}$ (I-band) and the R-band always deviated
by more than $1.8\,{\rm mag}$. This could partly be explained by
strong ${\rm H}_{\alpha}$ line emission, which falls within the
R-band. However, one has also to consider that the
UV-amplitude is poorly constrained (we only have a lower limit) which
limits the validity of the fitting procedure. Departure from
blackbody is not unexpected. For example, from Table~2 of
\citet{deJager1986}, we can similarly deduce non-blackbody emission
for a flare on the K6V dwarf BY Draconis.

The BUSCA 2005 data reveal another flare on 2M1707+64 at the last data
point on the morning of 2005.06.14. For this second flare we find (lower
limits to the) amplitudes of $3.70$, $0.94$, $0.77$ and $0.16$ magnitudes in
the UV, G, R and I-bands respectively. Although at the very end of
observations, the size of the amplitudes and the absence of variations
in the reference stars rule out an observational artifact.
Again, the model fits with the lowest $f$-value show the R-band
amplitude to deviate by more than $0.5\,{\rm mag}$ while the other three bands
can be reproduced to better than $0.1\,{\rm mag}$.

In total we obtained $22.3\,{\rm hr}$ of BUSCA data in 2002 and 2005
on 2M1707+64 which revealed two flares. With a total recorded flare
duration of $1.1\,{\rm hr}$ this implies a probability that the object
is flaring at any one time to be $0.049$. Note, however, that
the flare continued after the end of observations on that morning, so
this is a lower limit (see Fig.\ref{figure_LC2002}). If we instead
just count flares, we deduce a flare rate of $0.090 \pm 0.063\,{\rm
hr}^{-1}$ where the error is obtained by counting (Poisson)
statistics.  If we also take into account the $15.2\,{\rm hr}$ of
Maidanak data whose sensitivity is restricted to strong flares
(approx.\ $\Delta m_{\rm I}>0.05$, compare
Table~\ref{table_FlareProperties}), we obtain a flare rate of $0.029$
($0.053\pm 0.038\,{\rm hr}^{-1}$).

%%% Local Variables: 
%%% mode: latex
%%% TeX-master: "Paper"
%%% End: 

\section{Periodic variability in 2M1707+64}\label{section_PeriodicVariability2M1707}

From the five nights of follow-up observations in June 2005 we detect
periodic variations with periods (obtained from Scargle periodograms)
in the individual nights ranging from $3.61$ to $3.64$ hours at a
$99.9\,\%$ confidence level. This confirms the period we claimed in
R06. The application of the CLEAN periodogram to the combined
BUSCA-Maidanak data of 2005 yields a period of $P=3.619\pm 0.020\,(\pm
0.015)\,{\rm hr}$. The first uncertainty estimate (0.020 hr) was
obtained by Monte-Carlo simulations which account for the time
sampling (see R06).  The second estimate is from the theoretical
expression for the width of peaks in a periodogram, $\Delta
P=P^{2}/(2T)$, where $P$ is the period and $T$ the total time span
covered by the observations. This is only valid for uniform time
sampling, an assumption which is strongly violated
here. The former result overcomes this, although still assumes the
signal to be sinusoidal with Gaussian noise.

\begin{figure}
\includegraphics[width=84mm]{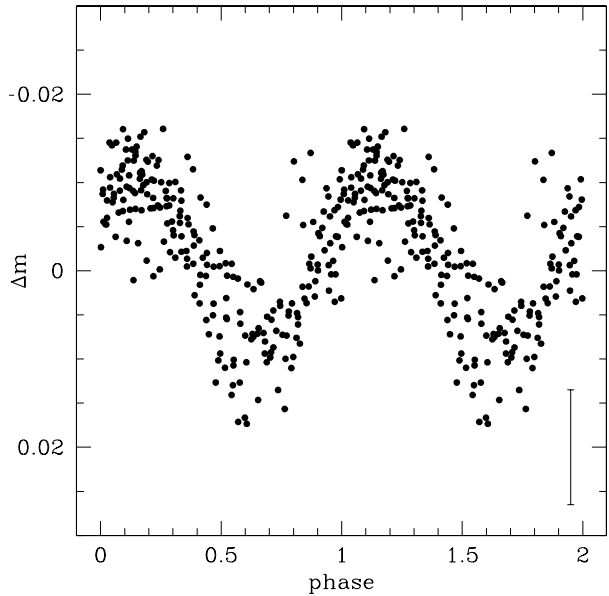}
\caption{2M1707+64: I-band light curve of the five nights between
2005.06.14 and 29 phased to a period of $3.619\,{\rm hr}$. A typical
error bar is also shown. Note that two cycles of the phased light
curve are plotted.}
\label{figure_2M1707PhasedLC}
\end{figure}

We can exploit the non-uniform time sampling in this case to
make a more robust estimate of the period uncertainty by constraining
a single sinusoidal light curve to fit the data on all nights. The two
consecutive BUSCA nights separated by several days from the three
Maidanak nights in 2005 (see Table~\ref{table_ObservationLog}) gives a
strong constraint on this fit. The best fit gives a period of
$3.619\,{\rm hr}$, the same as found with the CLEAN periodogram. To
estimate the uncertainty in this, we fix the phase of the sinewave at
2005.06.27 00:00UT ($=t_0$) and vary its period until we achieve a
phase shift of $\pi$ over the 13 day difference between the 14/15 June
observations and $t_0$.  The difference between this period and the
nominal one is 0.015\,hr; we (conservatively) define this to be the
uncertainty. It cannot be much larger, because if we further change
the period until we reach a phase shift of $2\pi$ (which still gives a
reasonable fit on 2005.06.14 \& 15), we now see a significant
deviation on 2005.06.24 \& 29. Hence, under the assumption that the
observed periodic variability on all five nights in June 2005 was
caused by the same star spot(s), and that differential rotation is
negligible \citep{Barnes2005}, we establish the rotation period (of
the spot system) to be $3.619\pm 0.015\,{\rm hr}$, an accuracy of
better than $1\,{\rm min}$.  Although the periodogram implies
a similar uncertainty, the sinewave fit is more
robust.\footnote{Mathematically the uncertainty derived from the
sinewave-fitting is $ \frac{1}{2} \frac{P}{T} \times P$, identical to
the theoretical expression for the periodogram. However, the
derivation of the periodogram expression depends on the (invalid)
assumption of uniform sampling. With the fitting method, in contrast,
we explicitly determine the quality of fit of the data to the model at
individual epochs for arbitrary phase shifts.}  The lifetime of
stellar surface spots can be much longer than two weeks
\citep{Strassmeier1992} so our assumption is reasonable. On the other
hand, the lifetime of spots on very low mass stars and brown dwarfs
has been little studied and may differ from G and K~stars. If the spot
distribution evolves significantly during the 2005 observations then
the fitting method is not valid. The data from the five nights phased
to $3.619\,{\rm hr}$ are shown in Fig.~\ref{figure_2M1707PhasedLC}.

%%% Local Variables: 
%%% mode: latex
%%% TeX-master: "Paper"
%%% End: 

\section{Flares on other objects}\label{section_FlaresOtherObjects}

From our 20 night BUSCA survey (R06), we found three further flares,
one in each of the following objects: 2M1344+77, 2M1546+37 and
2M1714+30. All three dwarfs have detectable quiescent $H_{\alpha}$
emission \citep{Gizis2000}. Amplitudes and further properties are
listed in Table~\ref{table_FlareProperties}. R06 detected variability
in the R-band in 2M1546+37 and 2M1714+30 at an amplitude of
$0.012\,{\rm mag}$ in both cases. A tentative period of $6.9\,{\rm
hr}$ was assigned to the latter in the R-band only.
\begin{table*}
\begin{minipage}{150mm}
\caption{Properties of the detected flares: spectral type, quiescent
$H_{\alpha}$ equivalent widths, the flares' amplitudes in the four
bands, an estimate to their duration and the number of recorded data points (samples)
for each flare. $H_{\alpha}$ EWs are taken from \citet{Gizis2000}.}
\label{table_FlareProperties}
\begin{tabular}{lllllllllc}
\hline
object  &  year  &  SpT  &  $H_{\alpha}$ EW [\AA]  &  $\Delta_{\rm UV}$  &  $\Delta_{\rm G}$
&  $\Delta_{\rm R}$  &  $\Delta_{\rm I}$  &  duration [hr]  &  samples\\
\hline
2M1707+64  &  2002  &  M9  &  9.8  &  $>6$  &  2.9  &  2.7  &  1.0  &  $>1.0$  &  4\\
2M1707+64  &  2005  &      &       &  $>3.7$  &  0.94  &  0.77  &  0.16  &  $>0.1$  &  1\\
\hline
2M1344+77  &  2002  &  M7  &  2.7  &  $>1.6$  &  0.094  &  0.047  &  0.007  &  $<0.5$  &  1\\
2M1546+37  &  2003  &  M7.5  &  10.9  &  $>1.7$  &  0.155  &  0.121  &  0.022  &  $<0.75$  &  1\\
2M1714+30  &  2003  &  M6.5  &  5.4  &  $>1.4$  &  0.075  &  0.031  &  0.007  &  $<0.5$  &  1\\
\hline
\end{tabular}
\end{minipage}
\end{table*}
In all three cases the blackbody model fits the observed amplitudes
relatively well and the following applies to all three flare
events. If we again consider those simulations with low $f$-values, we
find two classes of results. One well reproduces I, R and G amplitudes
with a strong deviation in the UV-band ($\approx 1.1\,{\rm mag}$) and
the other fits the I, R and UV well with a deviating G-band amplitude
($\approx 0.2\,{\rm mag}$). Since the UV-amplitude is not as
well determined as the other three (see
Sect.~\ref{section_Flares2M1707+64}), the former seems to be more
likely and thus indicates relatively good agreement with a blackbody
emission. However, various emission lines have been observed in flare
spectra in both bands \citep{Eason1992} and could thus modify the
continuum emission (that itself also could be non-blackbody) such that
it deviates from a blackbody. This could explain any residual
deviations.

%%% Local Variables: 
%%% mode: latex
%%% TeX-master: "Paper"
%%% End: 

\section{Conclusions}

From $227.6$ hours of multiband monitoring of 19 M~dwarfs
($218.0\,{\rm hr}$ with BUSCA in 2002 (R06) and $9.6\,{\rm hr}$ with
BUSCA in 2005), we detected five flares on 4 different targets. From
this we infer a flare probability at any one time of 0.013 and a flare
rate of $0.022\pm 0.010\,{\rm hr}^{-1}$ for field M2--M9
dwarfs. Although the observation time per object is limited (less than
14 hours per object, except for the follow up observations of
2M1707+64), we conclude that flare activity is not rare in
old (field) mid to late M~dwarfs and might even be present in most of
these objects.

We discovered a massive flare on the M9~dwarf 2M1707+64 with a
UV~amplitude of more than 6~mag. This object, which is either a very
low mass star or brown dwarf, shows a repeat of flare activity at a
rate of $0.090\pm 0.063\,{\rm hr}^{-1}$ (probability of flaring
$0.049$).  With both this and another flare on the same object we
found deviations of the flares' 4-band emission from that expected for
a blackbody. In particular, the R-band always fit poorly,
which is consistent with the presence of strong $H_{\alpha}$
line emission.

We also improved the determination of the rotation period of
2M1707+64, deriving a value of $3.619\pm 0.015\,{\rm hr}$, under the
assumption that the observed spot was stable over 16 days of
observations.

We found three further flares on the late~M dwarfs 2M1344+77,
2M1546+37 and 2M1714+30. Their amplitudes and energy release were much
smaller than in the case of 2M1707+64 and their emission closer
resemble that of a blackbody. The occurrence of flares and cool
surface spots strongly suggests the existence of magnetic fields in
UCDs, even though the exact creation mechanism remains enigmatic.

%%% Local Variables: 
%%% mode: latex
%%% TeX-master: "Paper"
%%% End: 

\vspace{1cm}
We are very grateful to Roland Gredel, the director of the Calar Alto
observatory, for a prompt and uncomplicated allocation of Directors
Discretionary Time in 2005. We also thank the Calar Alto staff for obtaining
these data and for their support during the 2002 and 2003 observing runs.

%\bibliographystyle{mn2ea}
%\bibliography{Bibliography}

\begin{thebibliography}{}

\bibitem[\protect\citeauthoryear{Allard, Hauschildt, Alexander, Tamanai \&
  Schweitzer}{Allard et~al.}{2001}]{Allard2001}
Allard F.,  Hauschildt P.~H.,  Alexander D.~R.,  Tamanai A.,    Schweitzer A.,
  2001, ApJ, 556, 357

\bibitem[\protect\citeauthoryear{Bailer-Jones \& Lamm}{Bailer-Jones \&
  Lamm}{2003}]{Bailer-Jones2003}
Bailer-Jones C.~A.~L.,  Lamm M.,  2003, MNRAS, 339, 477

\bibitem[\protect\citeauthoryear{Barnes, Collier~Cameron, Donati, James,
  Marsden \& Petit}{Barnes et~al.}{2005}]{Barnes2005}
Barnes J.~R.,  Collier~Cameron A.,  Donati J.-F.,  James D.~J.,  Marsden S.~C.,
     Petit P.,  2005, MNRAS, 357, L1

\bibitem[\protect\citeauthoryear{Barrado~y Navasc\'ues, Zapatero~Osorio,
  Mart\'in, B\'ejar, Rebolo \& Mundt}{Barrado~y Navasc\'ues
  et~al.}{2002}]{Barrado2002}
Barrado~y Navasc\'ues D.,  Zapatero~Osorio M.~R.,  Mart\'in E.~L.,  B\'ejar
  V.~J.~S.,  Rebolo R.,    Mundt R.,  2002, A\&A, 393, L85

\bibitem[\protect\citeauthoryear{Batyrshinova \& Ibragimov}{Batyrshinova \& Ibragimov}{2001}]{Batyrshinova2001}
Batyrshinova V.~M., Ibragimov M.~A.,  2001, AL, 27(1), 29

\bibitem[\protect\citeauthoryear{de Jager, Heise, Avgoloupis, Cutispoto,
  Kieboom, Herr, Landini, Langerwerff et~al.}{de~Jager
  et~al.}{1986}]{deJager1986}
de Jager C.,  Heise J.,  Avgoloupis S.,  Cutispoto G.,  Kieboom K.,  Herr
  R.~B.,  Landini M.,  Langerwerff A.~F. et~al.  1986,
  A\&A, 156, 95

\bibitem[\protect\citeauthoryear{Delfosse, Forveille, Perrier \&
  Mayor}{Delfosse et~al.}{1998}]{Delfosse1998}
Delfosse X.,  Forveille T.,  Perrier C.,    Mayor M.,  1998, A\&A, 331, 581

\bibitem[\protect\citeauthoryear{Eason, Giampapa, Radick, Worden \& Hege}{Eason
  et~al.}{1992}]{Eason1992}
Eason E.~L.~E.,  Giampapa M.~S.,  Radick R.~R.,  Worden S.~P.,    Hege E.~K.,
  1992, AJ, 104

\bibitem[\protect\citeauthoryear{Fuhrmeister \& Schmitt}{Fuhrmeister \&
  Schmitt}{2004}]{Fuhrmeister2004}
Fuhrmeister B.,  Schmitt J.~H.~M.~M.,  2004, A\&A, 420, 1079

\bibitem[\protect\citeauthoryear{Fuhrmeister, Schmitt \&
  Hauschildt}{Fuhrmeister et~al.}{2005}]{Fuhrmeister2005}
Fuhrmeister B.,  Schmitt J.~H.~M.~M.,    Hauschildt P.~H.,  2005, A\&A, 436,
  677

\bibitem[\protect\citeauthoryear{Gizis, Monet, Reid, Kirkpatrick, Liebert \&
  Williams}{Gizis et~al.}{2000}]{Gizis2000}
Gizis J.~E.,  Monet D.~G.,  Reid I.~N.,  Kirkpatrick J.~D.,  Liebert J.,
  Williams R.~J.,  2000, AJ, 120, 1085

\bibitem[\protect\citeauthoryear{Koen}{Koen}{2005a}]{Koen2005b}
Koen C.,  2005a, MNRAS, 360, 1132

\bibitem[\protect\citeauthoryear{Koen}{Koen}{2005b}]{Koen2005a}
Koen C.,  2005b, MNRAS, 357, 1151

\bibitem[\protect\citeauthoryear{Liebert, Kirkpatrick, Cruz, Reid, Burgasser,
  Tinney \& Gizis}{Liebert et~al.}{2003}]{Liebert2003}
Liebert J.,  Kirkpatrick J.~D.,  Cruz K.~L.,  Reid I.~N.,  Burgasser A.,
  Tinney C.~G.,    Gizis J.~E.,  2003, AJ, 125, 343

\bibitem[\protect\citeauthoryear{Liebert, Kirkpatrick, Reid \& Fisher}{Liebert
  et~al.}{1999}]{Liebert1999}
Liebert J.,  Kirkpatrick J.~D.,  Reid I.~N.,    Fisher M.~D.,  1999, ApJ, 519,
  345

\bibitem[{Mart\'in {et~al.}(1999)Mart\'in, Delfosse, Basri, Goldman, Forveille,
  \& Zapatero~Osorio}]{Martin1999}
Mart\'in, E.~L., Delfosse, X., Basri, G., {et~al.} 1999, AJ, 118, 2466

\bibitem[\protect\citeauthoryear{Mohanty \& Basri}{Mohanty \&
  Basri}{2003}]{Mohanty2003}
Mohanty S.,  Basri G.,  2003, ApJ, 583, 451

\bibitem[\protect\citeauthoryear{Rockenfeller, Bailer-Jones \&
  Mundt}{Rockenfeller et~al.}{2006}]{Rockenfeller2006a}
Rockenfeller B.,  Bailer-Jones C.~A.~L.,    Mundt R.,  2006, accepted by A\&A

\bibitem[\protect\citeauthoryear{Scholz \& Eisl\"offel}{Scholz \&
  Eisl\"offel}{2005}]{Scholz2005}
Scholz A.,  Eisl\"offel J.,  2005, A\&A, 429, 1007

\bibitem[\protect\citeauthoryear{Stepanov, Fuerst, Krueger, Hildebrandt, Barwig
  \& Schmitt}{Stepanov et~al.}{1995}]{Stepanov1995}
Stepanov A.~V.,  Fuerst E.,  Krueger A.,  Hildebrandt J.,  Barwig H.,
  Schmitt J.,  1995, A\&A, 299, 739

\bibitem[\protect\citeauthoryear{Strassmeier \& Bopp}{Strassmeier \&
  Bopp}{1992}]{Strassmeier1992}
Strassmeier K.~G.,  Bopp B.~W.,  1992, A\&A, 259, 183

\end{thebibliography}

\end{document}